# Proposed experiments for detecting contextual hidden variables

22.05.2025

Konstantinos Papatryfonos[1], Louis Vervoort[2]

[1] *Institute of Electronics, Microelectronics and Nanotechnology (IEMN), UMR CNRS 8520, University of Lille, Avenue Poincare, 59650 Villeneuve d'Ascq, France, konstantinos.papatryfonos@iemn.fr*
[2] *Higher School of Economics, School of Philosophy, Moscow, Russian Federation, lvervoort@hse.ru*

**Abstract**. We propose two quantum experiments – modified Bell tests – that could detect contextual hidden variables underlying quantum mechanics. The experiments are inspired by hydrodynamic pilot-wave systems that mimic a wide range of quantum effects and exhibit a classical analog of contextuality. To justify the experiments, we show that contextual hidden variables are inevitable and 'physics as usual' if a unification between quantum mechanics and general relativity is possible. Accordingly, contextual theories can bypass Bell's theorem in a way that is both local and non-conspiratorial. We end with a note on the relevance of exploratory experiments in the foundations of quantum physics.

## 1. Introduction

Since nearly two decades, physicists have been studying hydrodynamic 'pilot-wave' systems – droplets walking on vibrating fluid films – that mimic a wide range of quantum properties. These fluid/quantum analogies include single-particle double-slit diffraction [1, 2], tunneling [3, 4], quantized orbits [5, 6], quantum corrals [7, 8], Hong-Ou-Mandel interference [9], supperradiance [10, 11, 12], Anderson localization [13], violation of a Bell inequality [14], as well as further recent developments summarized in [15, 16]. After careful experimental and theoretical analysis, it has become clear that the observed quantum-like behavior can be explained by the pilot-wave that accompanies the droplets. Specifically, as the droplets bounce on the fluid surface, they generate a wave field that resonantly guides their horizontal motion, creating a wave-particle 'duality' involving a droplet and its pilot-wave. As a consequence, this macroscopic wave-particle system exhibits (a type of) contextuality: the wave and hence the movement of the droplet is influenced by the whole experimental set-up, including the geometry of the bath boundaries at a far distance from the droplet. This contextuality explains for instance the violation of a static Bell inequality in this system, as recently reported [14]. Indeed, in this static Bell test the left (right) droplet 'feels' the detector at the right (left) side of the experiment. Of course, contextuality is usually a term restricted to the quantum realm [17, 18, 19], where, however, it seems far from having disclosed all of its mystery. There is an intimate link between quantum contextuality and nonlocality[1]. As Mermin puts it: "The [Bell-Kochen-Specker] theorems establish that in a hidden-variables theory the values assigned even to a set of mutually commuting observables must depend on the manner in which they are measured—a fact that Bohr could have told us long ago (although he would have disapproved of the whole undertaking). And Bell's Theorem establishes that the value assigned to an observable must depend on the complete experimental arrangement under which it is measured even when two arrangements differ only far from the region in which the value is ascertained" [18]. In the droplet system, 'the value assigned to an observable [or property]' does indeed 'depend on the complete experimental arrangement under which it is measured' (see e.g. [14]). So, there is at least a partial analogy.

---
[1] Nonlocality is often believed to be a special case of contextuality, e.g. [19].

In this article we use the fluid-mechanical system as a source of inspiration for proposing new Bell-type quantum experiments that could lead to the detection of contextual hidden variables. The first experiment is inspired by a remarkable feature of the pilot-wave dynamics described in the next Section. Specifically, we perform simulations using the model developed by Papatryfonos et al. [14] over a new parameter range, and show that the Bell correlation function depends on the range of the 'hidden variables' ($\Delta\lambda$), the role of which is taken in this hydrodynamic system by the initial droplet positions. Based on these results, we propose a test that, if successful, could detect analogous degrees of freedom underlying the quantum realm. Experiments of the second type have already been done [20, 21], but we believe the remarkable fluid/quantum analogies warrant renewed interest. The hydrodynamic pilot-wave is much faster than the droplet speed, and could thus explain Bell-inequality violation also in dynamic Bell experiments on such droplets, as long as the rate of changing the barrier setting is slower than the time it takes the waves to travel through the system. Analogously, in photon experiments Bell-nonlocality could, in principle, be explained by such a faster-than-light pilot-wave. Notice that there is a key difference between the two types of experiments: the second experiment explores the existence of faster-than-light communication, while the first one tests the existence of contextual properties that could be strictly local.

Naturally, the whole enterprise of searching for hidden variables remains speculative (but see Section 4 for a justification), and we cannot offer a precise numerical prediction to be tested. But we propose testing qualitative trends grounded on well-understood physics, which could discriminate between quantum and sub-quantum behavior, i.e. behavior that can only be explained by a more fundamental theory than quantum mechanics. Hence, this research may be seen as part of a larger effort of discovering additional variables that can explain quantum theory. Such variables are denied existence by the orthodox Copenhagen interpretation, and are believed to be vastly constrained by no-go theorems, such as Bell's theorem. For the time being, no widely accepted candidates exist for such hidden properties. However, there is a qualitative analogy between the hydrodynamic pilot-wave and the pilot-wave of de Broglie's 'double-solution' theory [22] (for a summary of de Broglie's evolving ideas, see [23] and references therein; for a comparison with the hydrodynamic systems, see [24]). As is well-known, de Broglie's seminal approach delivered the key idea leading to Schrödinger's wave mechanics *and* the de Broglie-Bohm theory. Thus, although one can remain agnostic as to the nature of the hidden variables, our proposed experiments could – broadly – be seen as an exploration of de Broglie's pilot-wave ideas. In a series of interesting articles, Drezet and Jamet have elaborated on de Broglie's double-solution program, which allows them to propose a detailed interpretation of nonlocality and other key quantum properties [23, 25, 26, 27]. We make a comparison between their view and our interpretation of contextuality and nonlocality in Section 4.

Furthermore, other researchers have proposed exploratory models from which, if finalized, quantum theory could be derived in some limit (e.g. [28, 29, 30, 31]). We submit that all these theories could be termed contextual, as explained in Section 4 – in agreement with e.g. Mermin's position cited above. Our main justification of the present effort is that such contextual variables might bypass no-go theorems in a local *and* non-conspiratorial way, as argued in Section 4. As a final justification of the present work, let us recall that great theoretical challenges remain in the foundations of physics, notably related to the unification of quantum mechanics and general relativity; the interpretation of quantum theory; the measurement problem; etc. If theory is stagnating, experiments may offer valuable insights, even if exploratory. Interestingly, the situation is similar in particle physics, where a relative consensus reigns as to the usefulness of open-ended experiments, experiments that are not necessarily guided by a finalized theory (e.g. in the context of dark matter and 'beyond Higgs' physics [32]).

The article is organized as follows. In Section 2 we describe the numerical Bell experiment that triggered the present study, as well as the model used for the current simulations. This investigation is done in a hydrodynamic Bell setting, in which correlated left and right droplets can 'tunnel' between two wells; their position in the inner or outer well corresponds to the binary Bell property being +1 or -1, as in [14]. In Section 2 we present the results of the calculations of the corresponding Bell correlation function as a function of the range of initial values, noticing a distinct difference with what one expects in quantum Bell experiments according to standard quantum theory. We then propose in Section 3 a modified Bell experiment that could detect whether an analogous contextual behavior exists in photon experiments. In this section we also present a second experiment that could detect unknown contextual properties. In Section 4 we give a further justification of the relevance of contextual sub-quantum variables, i.e. yet unknown variables depending on the detector/analyzer angles in Bell experiments. We argue that contextual hidden-variable theories are to be expected, and 'physics as usual', if one starts from innocuous assumptions – that quantum mechanics and relativity theory can be unified, or that 'everything is physical'. Section 5 concludes.

## 2. Hydrodynamic Bell test, old and new

### 2.1. Previous work

Recall, first, that a Bell test is 'system-independent': it can in principle be done on *any* system, microscopic or macroscopic, that is assumed to satisfy the Bell conditions, i.e. the conditions needed to derive a Bell inequality. In a wider sense, such a test can be done on systems consisting of two separated subsystems, and on which a binary property X, whatever X may be, is measured as a function of detector settings ('$\alpha$' on the left, '$\beta$' on the right). If the system is local and classical, it is expected to satisfy a Bell inequality. Here, we look at a system that is only local if considered naïvely, namely if one does not know of the delocalized nature of the pilot-wave that accompanies the sub-systems — the bouncing droplets in our case[2]. In principle, the hydrodynamic test described below could be done in a dynamic setting that imposes locality, as proposed in [33]. In this Section, however, we present results for a static Bell setting, which is numerically more easily tractable, and which suffices to display intriguing contextual dynamics of the droplet system.

---

[2] Physicists needed years of investigation to understand the range and detailed dynamics of the pilot-wave in the droplet system, and to explain how it could mimic quantum behavior. If one would not know of the wave and only consider the droplets, the system would be deemed local.

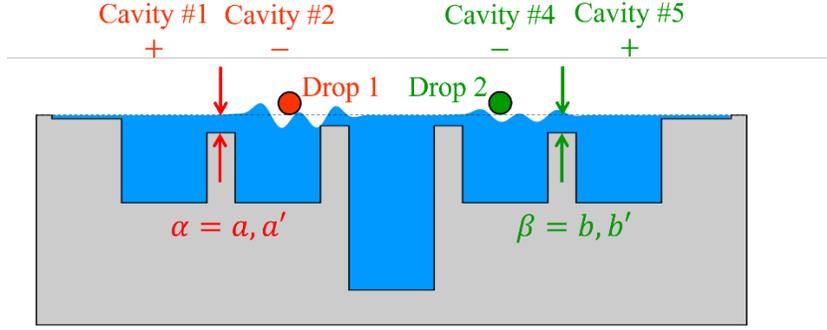

Figure 1: Schematic of a hydrodynamic Bell test (adapted from Papatryfonos et al. [14]). The system comprises two droplets (red and green) walking on the surface of a vibrating liquid bath (blue) over a solid substrate (grey). Each droplet is confined within a subsystem consisting of a pair of cavities separated by barriers, across which the droplets tunnel unpredictably. The tunneling rate is influenced by the barrier depths, denoted $\alpha$ and $\beta$, which can take the values (a, a') on the left and (b, b') on the right. The symbols (+) and (-) indicate whether a cavity corresponds to the ground (-) or excited (+) energy state of the respective droplet within its subsystem.

A schematic representation of the numerical Bell-type experiment that inspired this study is shown in Figure 1. A detailed description of the system and of its potential application as a platform for exploring Bell correlations can be found in [14]. The system consists of two droplets walking on the surface of a vibrating liquid bath, over a solid substrate. Each droplet is confined within a subsystem composed of a pair of cavities separated by barriers; the droplets may unpredictably tunnel across these barriers. The tunneling rate is influenced by the barrier depths (playing the role of detector settings in our Bell experiment), denoted $\alpha$ and $\beta$, which can take the values (a, a') on the left and (b, b') on the right.

The two subsystems, labeled A and B, each contain a single droplet which generates its own waves after each bounce on the surface. The droplets remain confined to a pair of identical cavities due to a pair of higher barriers separating the central cavity from the 4 outer cavities. While these central barriers are high enough to prevent the drops from penetrating the middle cavity, they are also low enough to allow the waves generated by the droplets to pass through, thus creating coupling between the two subsystems. The strength of this coupling depends on the central cavity's geometry; increasing its depth $d_c$ enhances the coupling, which enables it to act as a nearly resonant transmission line [34], while altering its width affects the coupling in a non-trivial manner [11].

Within each subsystem, the lower energy state (−) corresponds to the cavity in which the droplet tends to spend most of its time, while the other cavity corresponds to the excited state (+). In general, the ground state may be either the inner or outer cavity, depending on the geometry of the system [11]. For the geometry studied here, the inner cavity is the ground state (−), and the outer cavity is the excited state (+), as shown in Fig. 1. Transitions between ground and excited states within the subsystems correspond to individual tunneling events, with rates determined by the depths of the submerged barriers. These barrier depths ($\alpha$ for subsystem A and $\beta$ for subsystem B) serve as the measurement settings in our Bell tests, as in [14]. The Bell measures $X_A$ and $X_B$ indicate whether the droplets are in the inner well ($X_j = -1$, j = A, B) or the outer well ($X_j = +1$, j = A, B) at the measurement time $T_m$. The properties $X_j$ are calculated via the coupled equations (2.4) – (2.6) involving the positions $x_j(t)$, allowing to determine $x_j(T_m)$ and hence $X_j$, j = A, B. One can then define a Bell experiment in which droplets are pair-wise generated above the central cavity, and move to their respective sides, where their positions $X_{A,B}$ are measured at $T_m$ as a function of $(\alpha, \beta)$. In real experiments, perhaps a slight impulse, for instance imparted by a fine airflow from a thin tube, might be needed to make the droplets move in opposite directions. In our simulations we let them simply start in the inner or outer cavities and calculate the subsequent resonant tunneling events.

In a local and classical system one expects that

$$|S(a, b, a', b')| \leq 2 \quad (2.1)$$

for any choice of measurement settings $(a, a', b, b')$, where

$$S = M(a, b) + M(a', b) + M(a, b') - M(a', b'). \quad (2.2)$$

Here $M(\alpha, \beta)$ is the average product

$$M(\alpha, \beta) = \sum_{X_A, X_B} X_A X_B P(X_A, X_B | \alpha, \beta). \quad (2.3)$$

To model the one-dimensional droplet motion we use the numerical method developed by Nachbin [34, 35] and adapted in [11, 14] in order to simulate the tunneling in the geometry depicted in Fig. 1. Each droplet generates waves and interacts with them according to Eqs. (2.4)–(2.6).

$$m\ddot{x}_j + cF(t)\dot{x}_j = -F(t)\frac{\partial \eta}{\partial x}(x_j(t), t) \quad (2.4)$$

$$\frac{\partial \eta}{\partial t} = \frac{\partial \phi}{\partial z} + 2\nu \frac{\partial^2 \eta}{\partial x^2} \quad (2.5)$$

$$\frac{\partial \phi}{\partial t} = -g(t)\eta + \frac{\sigma}{\rho}\frac{\partial^2 \eta}{\partial x^2} + 2\nu \frac{\partial^2 \phi}{\partial x^2} - \sum_{j=1,2} \frac{P_d(x - x_j(t))}{\rho} \quad (2.6)$$

In some detail, the wave model is based on a weakly viscous quasipotential theory [36], which was recently extended to capture vertical dynamics, including the behaviour of the lubricating air layer during the rebound droplet dynamics [37]. In Equations (2.4)-(2.6), *η(x,t)* represents the wave elevation and *ϕ(x,z,t)* the velocity potential, which satisfies Laplace's equation in the fluid domain. The fluid is characterized by its density *ρ*, surface tension *σ*, and kinematic viscosity *ν*. At the free surface (*z = 0*), the wave dynamics are governed by Eqs. (2.5) and (2.6), with the pressure term *P_d* in Eq. (2.6) representing the influence of the droplet. Specifically, acting as a wave maker, the droplet applies pressure at its instantaneous position, *x(t)*, during contact with the surface. The droplet's horizontal motion is coupled to this wave system through Eq. (2.4), in which the propulsive wave force imparted to the droplet during surface contact, *F(t)*, also contributes to the damping coefficient [14].

The main result of [14] is that the 'static' Bell inequality (2.1) can be violated in this system (up to S = 2.49 ± 0.04) for a narrow range of parameter values. This effect is attributed to the delocalized correlation between the droplets, which is mediated through their common pilot-wave.

For instance, when one of the droplets tunnels to its excited state, the probability that also the second droplet does so increases substantially. Thus, through its wave-mediated interaction with its partner, each droplet is affected by the barrier depth of the distant station: a kind of classical contextuality.

*2.2. Present study*

In the new investigation reported here, we examine whether the correlation function $M(\alpha,\alpha)$ [Eq. (2.3) with $\beta = \alpha$] of the Bell inequality (2.1) is influenced by variations in the initial positions of the droplets, which are randomly chosen in a precision / uncertainty range $\Delta\lambda$. As above, we record the position of the droplets at the measurement time $T_m$ and calculate $M(\alpha,\alpha)$ for a given value of $\alpha$ according to Eqs. (2.3)–(2.6). More precisely, in each run, the starting position of a left droplet is assigned a random value within the interval $\Delta\lambda$ located at the center of the outer left cavity; the initial position of the right droplet is chosen independently in the symmetric interval $\Delta\lambda$ around the center of the outer right cavity. For each $\Delta\lambda$ and $T_m$, the simulation is repeated multiple times until the relative error in the running average of $M(\alpha,\alpha)$ falls below 3%, as in [14].

System parameters are chosen to represent a fluid bath with a density of 0.95 g/cm$^3$, viscosity 16 cS, and surface tension 20.9 dyn/cm, vibrating vertically in a sinusoidal manner with a peak amplitude $A_0$ and frequency $\omega/2\pi = 80$ Hz. This corresponds to a vertical acceleration amplitude of $A_0\omega^2 = 4.23$ g, which equals to 90 % of the Faraday threshold of the central cavity ($\gamma_F = 4.69$ g). The resonant bouncing of a droplet at the Faraday frequency generates a quasi-monochromatic damped wave pattern with a Faraday wavelength of $\lambda_F = 4.75$ mm. Each of the four outer cavities has a fixed length L of 1.0 cm ($\approx 2.1\ \lambda_F$) and a depth of 0.5 cm. The central cavity, coupling the right and left subsystems, has a length of 0.4 cm. Each barrier has a width of 0.4 cm, while the two barriers separating the central cavity from the subsystems have a depth of 0.045 cm. The barrier depth $\alpha$ (cf. Fig. 1), corresponding to the detector variable, has a value of 0.099 cm.

Figures 2 and 3 present the main results of this simulation, highlighting the dependence of the Bell correlation on $\Delta\lambda$. Figure 2 shows $M(\alpha,\alpha)$ for $\alpha = 0.099$ cm and $T_m = 1000$ Faraday periods ($T_F$); $T_F = 0.025$ s. Figure 3 corresponds to $T_m = 400\ T_F$. In both cases, we observe a strong dependence of $M(\alpha,\alpha)$ on $\Delta\lambda$. Moreover, the trend of this dependence varies significantly with the measurement time, as seen by comparing Figure 2 ($T_m = 1000\ T_F$) and Figure 3 ($T_m = 400\ T_F$), again a consequence of the complex dynamics governing this system. Since the simulations are computationally demanding, we show only a limited number of high-precision numerical results, but we have verified that these results are well representative of the dynamics of this system.

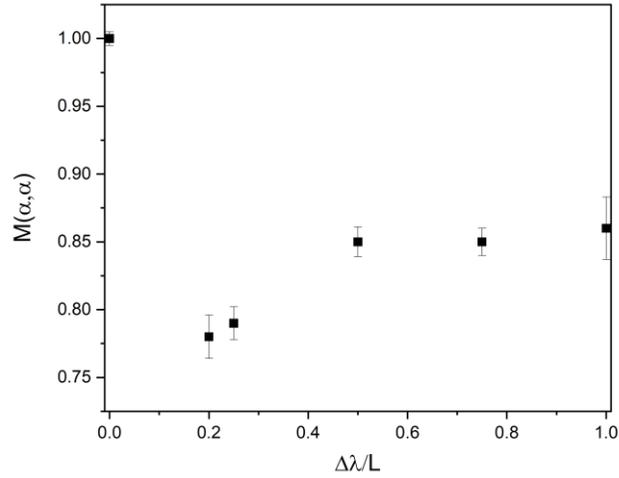

Figure 2: Bell correlation function M(α,α) (α = 0.099 cm) as a function of the range Δλ of the initial droplet positions, for measurement time $T_m = 1000\ T_F$. Error bars are indicated.

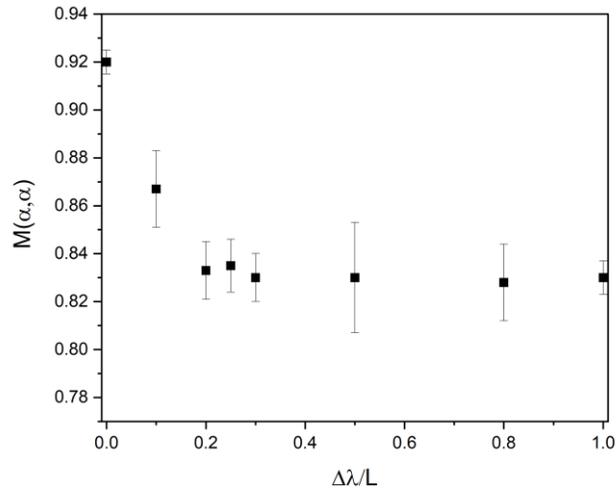

Figure 3. Bell correlation function M(α,α) (α = 0.099 cm) as a function of the range Δλ of the initial droplet positions, for measurement time $T_m = 400\ T_F$. Error bars are indicated.

Figure 4 presents a variant of the simulations, where the initial positions of each droplet pair are now perfectly symmetrical. These positions are determined by selecting a random number within the interval Δλ. The position of both droplets is then defined by using this random number as the distance from the center of their respective outer cavities.

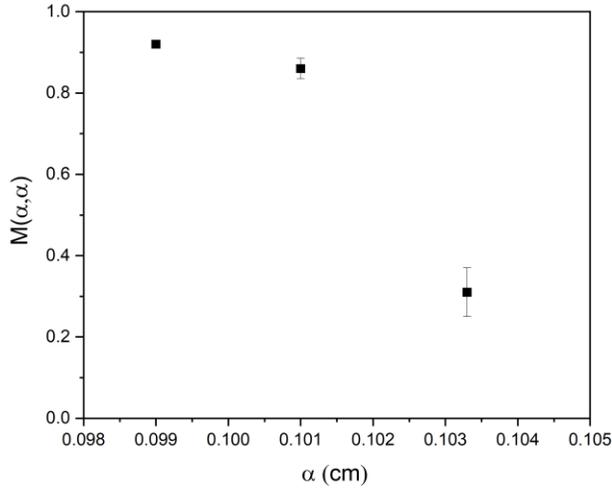

Figure 4. Bell correlation function $M(\alpha,\alpha)$ as a function of $\alpha$ for $\Delta\lambda/L = 1$ and for measurement time $T_m = 1200\ T_F$. Error bars are indicated.

In the next section we exploit these findings to propose a new quantum experiment.

## 3. Proposed quantum experiments

### 3.1. Cooled Bell experiment

As shown in Figs. 2-4, the correlation function $M(\alpha,\alpha)$ is in general not equal to 1, even when, for each droplet pair, the left and right droplet start from exactly symmetrical positions (randomly chosen in an interval $\Delta\lambda$), as in Fig. 4 or in Fig. 3 for $\Delta\lambda = 0$. Since the whole system is then exactly symmetrical, one expects that the left and right droplet end up at the same final position at the measurement time, so that $M(\alpha,\alpha) = 1$. In our simulations there appears to be a clear dependence of $M(\alpha,\alpha)$ on $\Delta\lambda$ (Figs. 2-3).

This deviation from $M(\alpha,\alpha) = 1$ can be attributed to the presence of tiny errors or uncertainties in the numerical values of the initial and subsequent positions, stemming from the inherent limitations of numerical calculations [11, 14, 38]. The nonlinear, chaotic nature of the hydrodynamic system enhances these tiny asymmetries in the course of movement. This, then, can explain why the left and right trajectories diverge, and why $M(\alpha,\alpha) \neq 1$ in Fig. 4 and when $\Delta\lambda = 0$ in Fig. 3. At the same time, there is a general trend of decreasing $M(\alpha,\alpha)$ when $\Delta\lambda$ increases. The fact that for $T_m = 1000\ T_F$ there is a dip in $M(\alpha,\alpha)$ at $\Delta\lambda/L = 0.2$ illustrates how subtle the dynamics is. The droplets require an average of approximately 70 Faraday periods to traverse the cavity once back and forth, so at $T_m = 1000\ T_F$ they had ample time to tunnel and follow diverging trajectories. But it remains somewhat intriguing that for a smaller $\Delta\lambda/L = 0.2$ the correlation is lower than for a larger $\Delta\lambda/L = 1$. This result shows that the relationship between $\Delta\lambda/L$ and correlation is not straightforward, indicative of the complex dynamics.

In sum, in the mathematically perfectly symmetrical case one expects:

$$M(\alpha,\alpha) \to 1 \text{ if } \Delta\lambda \to 0, \forall \alpha. \tag{3.1}$$

Even if the droplets of a pair do not start exactly at symmetrical positions, (3.1) still should hold. In the real pilot-wave system (3.1) is approximately the case; the relevant observation is that there is a clear dependence of $M(\alpha,\alpha)$ on $\Delta\lambda$.

In view of the rich analogies between the walkers and quantum systems, these findings in the walker system suggest to explore whether a similar dependence of the Bell correlation $M(\alpha,\alpha)$ might exist in the quantum realm. This suggests following experiment: *to restrict or vary the domain of hypothetical hidden variables of a photon pair in a Bell experiment*, notably the variables describing the emission from the source (as in the walker analogy), *or* the variables related to the detectors. As far as we know, the most practical approach to verify whether a dependence of type (3.1) exists is to cool the photon source, and/or the detectors, to near absolute zero and subsequently slowly increase their temperatures, and to test whether there is a dependence between M and T, the temperature of source and/or detectors, a dependence presumably of the form:

$$M(\alpha,\alpha) \to 1 \text{ if } T \to 0, \forall \alpha. \tag{3.2}$$

In quantum mechanics, as long as cooling does not change the singlet state, one has $M(\alpha,\alpha) = 1 \forall \alpha$ and $\forall$ T. Any deviation from M = 1 (not explained by quantum mechanics) is indirect support for the existence of hidden variables. Note that a restriction of the possible values of hidden variables is also a key idea of [39] to test superdeterminism. According to the convention we adopt here, superdeterministic variables are an instance of contextual hidden variables: they violate the same property, namely statistical independence, as recalled in Section 4.

Fig. 4 suggests another quantum experiment: to measure $M(\alpha, \alpha)$ in a Bell set-up as a function of $\alpha$. Again, deviations of M = 1 not explained by quantum mechanics are indirect support for the existence of hidden variables. We did not do a systematic study of the dynamic behaviour of the droplet system as a function of $\alpha$ (beyond the simulations in Fig. 4), so we leave this here as a line of further research.

### *3.2. Fast-switching Bell experiment*

As noted above, there is an intriguing analogy between the hydrodynamic pilot-wave system and de Broglie's double-solution theory [22], in which quantum particles are accompanied by a real wave. Although speculative, several researchers have started exploring the idea that quantum nonlocality could involve a detectable superluminal signal – possibly mediated by an advanced wave[3]. More precisely, they have tested whether the violation of a Bell inequality in dynamic photon experiments, in which a space-like separation between the left and right detection events is imposed, could be explained by a superluminal interaction – or wave – exchanged by the particles [20, 21]. These Bell experiments are done over very large distances and involve fast switching of the detector angles. Assuming that the particles interact with a superluminal speed $V_{SL}$, the distance and switching speed can be chosen large enough so as to separate the left and right detection events (with respect to $V_{SL} > c$, where c is the speed of light). If the violation of the Bell-inequality would break down starting from a given combination of distance and switching speed, this would then be evidence for an exchange mechanism at speed $V_{SL}$. Until now, no break-down has been seen in the Bell-inequality violation up to speeds $V_{SL} = 10^4$ times the light velocity [20, 21].

It is remarkable that the droplet system offers a real physical implementation of de Broglie's idea of an advanced wave accompanying particles. In the droplet system the advanced wave has a velocity about an order of magnitude larger than the droplet velocity. In view of the many fluid/quantum analogies this system exhibits, we suggest that it makes sense to continue the experimental series of refs. [20, 21] using larger distances and/or faster switching, so that higher velocities can be probed. Since the last experiments, technological progress allows now to probe these higher velocities[4]. Note that in the hydrodynamic analog discussed in Section 2, one could in principle also perform dynamic Bell tests [33]. If one would do such a test with dynamic settings that separate the left and right detection events *with respect to the droplet speed but not with respect to the wave speed*, then one could violate the corresponding Bell inequality in this system, due to the 'nonlocal', i.e. 'faster-than-the-particle', signal exchanged via the pilot-wave, just as in the static test.

## 4. Contextual hidden variables justified

Bell's theorem shows that hidden variable theories (HVTs) reproducing quantum mechanics must violate at least one of the assumptions Bell made to derive his inequality. HVTs could in principle be non-local, i.e. invoke communication at superluminal velocity, as would be tested in the 'fast-switching' experiment above. Of course, such a non-local HVT is in tension with relativity theory, which makes it, understandably, unpopular in the physics community. Still, there is discussion about whether tachyonic fields can exist (e.g. [40]). In view of this discussion and of the quantum / fluid analogy described above, we believe that more experiments would make sense. Another option to bypass Bell's theorem are contextual HVTs, which violate the Bell condition that is often termed 'measurement independence', 'statistical independence', or 'freedom-of-choice'. This requirement can be expressed simply as a condition of probabilistic independence between the HVs and the detector variables (a, b):

$$P(\lambda|a,b) = P(\lambda). \tag{4.1}$$

---

[3] According to the standard interpretation, superluminal signals would imply retro-causality, which is usually considered problematic. At the same time, there is some debate about the possible existence of tachyons. For a possible solution, see [23], commented on below.

[4] Personal communication by N. Gisin.

By definition, we will *generally* call HVTs that violate this condition 'contextual': they possess additional degrees of freedom that depend on the 'context'; more precisely, the detectors and their properties (a, b). One will notice that there is at least a superficial analogy with the walker system, as explained in Sections 1 and 2. Note also that, in principle, nonlocal, i.e. superluminal, influences could lead to violation of (4.1), for instance when there is a nonlocal signal from the detectors to the particles. Usually violation of (4.1) is deemed more interesting when it does not involve superluminal influences.

Now, such contextual HVTs are often believed to have utterly unpleasant features. Many assume that violation of (4.1) is tantamount to a cosmic conspiracy, violation of free will, superdeterminism or nonlocality – or perhaps a combination of these. The underlying idea (for the first three culprits) is that (particle) properties $\lambda$ cannot be correlated with the free or random choice events (a, b), unless something grotesque is going on. But we will argue here that this widespread interpretation is premature. For instance, proponents of superdeterminism note that (4.1) could be violated in a local manner if there are common causes[5] between $\lambda$ and (a,b) – we call these common-cause variables '$\lambda$*' in the following. The variables of some ultimate Theory-of-Everything (ToE) unifying quantum mechanics and general relativity and describing the Big Bang could provide these common causes (we indicate this ab-initio ToE as $\lambda$*-ToE or $\lambda$*-HVT), following [41]. The first problem with this interpretation is that (4.1) is utterly plausible *at least for macroscopic properties*: most macroscopic systems are independent and even if they would initially be correlated, the correlations are usually washed-out over time; so Bell's assumption of (4.1) is a priori entirely intuitive[6]. But when probing fundamental properties of elementary particles as electrons or photons, the situation may be different (some arguments in [33, 42]). The second problem with superdeterminism is that the mentioned ab-initio HVs surely are intractable: no-one believes that it would be possible to construct a ToE, based on such $\lambda$*, allowing one to calculate literally everything, including Alice's choice of her next detector setting in her next Bell experiment. This 'unphysical' nature of superdeterminism is likely the main reason why it has garnered relatively few proponents in the past (see discussions of superdeterminism in e.g. [23, 42, 43, 44, 45, 30, 46, 47]).

However, recent work suggests that (4.1) can be violated while avoiding the incalculability of hard-core superdeterminism (we follow Nikolaev and Vervoort [41]). To see how, we only need to make the superdeterministic scheme just a little more sophisticated. In the hard-core superdeterministic scheme all quantum variables as (x, y) and also the $\lambda$ in (4.1) are supposed to be explained/caused by the $\lambda$*-variables. Let us write this scheme of reduction as follows (the arrow can be read as 'explains'):

$$\lambda^* \rightarrow x, y, \lambda; \quad \text{or, on the level of theories: } \lambda^*\text{-HVT} \rightarrow \text{QM.} \tag{4.2}$$

---

[5] According to the standard interpretation of statistical dependence, violation of (4.1) is due to either direct causality between $\lambda$ and (a, b), or due to common causes that influence both $\lambda$ and (a, b) (or both). Since in dynamic Bell experiments direct causality seems, a priori, excluded, the remaining option are common causes between $\lambda$ and (a, b). We use causal language here, even if this is in principle not necessary: we could use the notions of functional or probabilistic dependence, which are mathematically defined.

[6] Needless to say, in the droplet system, violation of (4.1) is not due to superdeterministic causes, cf. Section 2.

A slightly more subtle scheme is this:

$$\lambda^* \to \lambda \to x, y; \quad \text{or, on the level of theories: } \lambda^*\text{-HVT} \to \lambda\text{-HVT} \to \text{QM}. \tag{4.3}$$

The detector variables (a, b) are part of all of the above theories; they are, obviously, part of the quantum description for instance. Expression (4.3) simply says that (we assume that) there is an intermediate level of description between the ultimate but elusive $\lambda^*$-ToE and quantum mechanics (QM), namely some *effective* theory involving variables $\lambda$ ($\lambda$-HVT). Note that such 3-step schemes of reduction are well-known in physics. One has for instance an analogous scheme between classical mechanics, statistical mechanics, and thermodynamics:

$$\text{classical mech.} \to \text{statistical mech.} \to \text{thermodynamics.} \tag{4.4}$$

In this analogy the role of the effective variables $\lambda$ is taken by Z, the partition function of statistical mechanics, which allows calculating all thermodynamic properties. Statistical mechanics itself is, *in principle*, explained by Newtonian mechanics and its classical variables ($\lambda^*$ in the analogy). (The 'in principle' here will become the crux of our interpretation.)

Returning to scheme (4.3), the first assumed reduction ($\lambda^*$-HVT $\to$ $\lambda$-HVT) implies that we may write, in probabilistic terms:

$$P(\lambda|a,b) = \sum_{\lambda^*} P(\lambda|a,b,\lambda^*) P(\lambda^*|a,b) \neq P(\lambda). \tag{4.5}$$

The first equality in (4.5) is a direct application of the law of total probability, and expresses that $\lambda^*$ explains/causes $\lambda$. The inequality in (4.5) follows from the fact that $\lambda$ and (a, b) are correlated by common causes: the $\lambda^*$. Similarly, the second reduction in (4.3) ($\lambda$-HVT $\to$ QM) immediately implies:

$$P(x,y|a,b) = \sum_{\lambda} P(x,y|a,b,\lambda) P(\lambda|a,b), \tag{4.6}$$

which can be interpreted as stating that the $\lambda$ explain quantum mechanics, in particular the Bell correlations.

Now, the crux of this model lies in the fact that in the reductive scheme (4.3) ($\lambda^*$-HVT $\to$ $\lambda$-HVT $\to$ QM) the second reduction may be calculable, *while the first may be real but hidden (incalculable)*. This is also precisely what happens in the analogous scheme of statistical mechanics (4.4). Indeed, constructing statistical mechanics *requires the assumption of classical particle properties $\lambda^*$, even if these properties are incalculable or intractable*; the dynamics of the individual particles involving these variables cannot be calculated. In a similar vein one may assume that there exist, *as a matter of principle*, variables $\lambda^*$ that explain the violation of (4.1), as in (4.5), while these variables will remain for ever unknown. At the same time, nothing prohibits the existence of an effective *contextual* $\lambda$-HVT that allows calculating the Bell correlation, as in (4.6), and therefore the whole of quantum mechanics, since Bell's no-go theorem is bypassed. In sum, in our interpretation the hard-core superdeterministic $\lambda^*$ exist, but this level of reality does not need to be the object of theory building; these variables may be forgotten by physicists – except for the fact that they offer an in-principle explanation for the violation of (4.1). What is in the focus of theory building are effective λ-HVTs, which can complete quantum mechanics, just as statistical physics can be built as an effective theory for mechanics and explain thermodynamics. Such λ-HVTs can be local, since Bell's no-go verdict is bypassed; they are contextual theories, by definition. More details about this interpretation can be found in [41].

However inaccessible, we believe that assuming the existence of the λ* is a most natural thing to do in physics: *such an assumption is simply tantamount to assuming that everything is ultimately physical* – which presumably few physicists would contest. If everything is ultimately physical, then even choice events that may initially appear random are so as well; and then it is only natural to assume that there exists, *in principle*, a λ*-ToE that describes the universe since the Big Bang, including human choices. But then ipso facto all physical events, including choice events, are correlated (deterministically or probabilistically) through the common causes λ*. Above, we just made one extra assumption of an intermediate effective level. Clearly, some might call our contextuality a hidden type of superdeterminism[7]; fine; but it is not of the type abhorred by Bell, since we do not need it to do physics: effective theories can be built.

If this hidden superdeterminism can just be interpreted as the very natural idea that 'everything is physical', seeing here conspiracy appears to be an anthropocentric bias. Again, physicists do not need to worry about the hidden level, inaccessible to physics; they can concentrate on constructing contextual theories. Interestingly, the early attempts of 't Hooft [28], [29] and Donadi and Hossenfelder [30] do seem to follow the contextual scheme, involving effective variables (see [41] for details). For instance, in 't Hooft's Cellular Automation model [29], the role of the λ* is taken by fast-fluctuating DOFs, evolving via permutations as in a deterministic automaton; but one does not have to know the details of this gigantic state space; one can integrate these variables out and construct an effective theory involving new effective HVs (the slow variables of the model). It is also not difficult to construct mathematical toy-models that recover the quantum correlation for a Bell experiment via violation of (4.1) (see [48] for a review).

Let us now compare our scheme to the double-solution theory of Drezet and Jamet [23, 25, 26, 27]. The authors show in [23] that de Broglie – Bohm theory, famously nonlocal, can be turned into a local theory by introducing two new variables (associated with any quantum system), namely an advanced and a delayed wave (rather, the anti-symmetrical half-difference '$u(x)$' of the two), evolving in four-dimensional spacetime rather than in configuration space. Thus, this theory would achieve de Broglie's double-solution program: it reproduces the essential features of quantum mechanics via local trajectory-guided wavelets, one delayed wavelet being causal (influencing the future), the other advanced wavelet being retrocausal. By applying their theory to a Bell experiment, the authors argue that in order to circumvent nonlocality, the theory has to be superdeterministic: in a nutshell, the advanced wave centered on the particle carries information to the past about the settings (a, b) of a Bell experiment; the initial *u*-field contains in advance information about what will happen to it in the future via the interaction with the settings. Now, the authors also argue that this superdeterminism is only apparent, arising only for 'causal observers', i.e. the usual causal perspective of the common mortals we are. If one takes a more objective, non-anthropocentric viewpoint, and accepts that nature is time-symmetric (as the fundamental laws are), then the retrocausal and hence superdeterministic nature of the theory would appear not fundamental ([23] p. 21).

---

[7] It was called 'soft superdeterminism' in reference [33].

Since this model offers detailed calculations and a comparison between different theories including quantum mechanics and Bohmian mechanics, it is instructive in showing how quantum nonlocality, retrocausality[8] and superdeterminism are intimately related. At least some of these appear as features of (interpretations of) the mathematics of the models rather than fundamental. Importantly, we submit that the interpretation of Drezet and Jamet fits well to the scheme discussed above (even if this scheme is in principle best applicable to theories unifying quantum mechanics and general relativity). The extra variables of the double-solution theory (essentially $u(x)$) are indeed *local and effective* variables that are correlated with (a, b), thus violating statistical independence (4.1) (e.g. [27] p. 28), thus contextual in our terminology.

Drezet and Jamet seem to privilege retrocausality, or more precisely, time-symmetry as being the fundamental ontological property, from which superdeterminism and nonlocality emerge as non-fundamental features. The future will tell whether this is the most productive interpretation, i.e. whether it has the greatest problem-solving capacity. But one can also adopt, it seems, an ontology-light position, a synthesis between our interpretation of Bell's theorem and Drezet and Jamet's interpretation of their double-solution theory as applied to a Bell experiment. Such a synthesis would be this: nonlocality, contextuality, superdeterminism and retrocausality are all sides of the same multifaceted coin, i.e. different conceptual interpretations of counterintuitive mathematical theories and theorems. This view would also resonate with Mermin's position cited in Section 1. Of course, on our view, contextuality and superdeterminism seem most fundamental, since essentially reducible to a truism, namely the assumption that everything is physical. As we argued above, assuming this we get superdeterminism and contextuality (almost) for free.

Our above argument, based on taking the harmless idea that 'everything is physical' seriously, can be considered from a slightly different angle. If quantum mechanics and general relativity theory can be unified, the most natural interpretation is that the variables of the unified theory will be local, since relativity is local by construction. Then they should be contextual, so that Bell's no-go theorem is bypassed.

## 5. Conclusion

We have proposed two modified Bell experiments that could test the existence of yet unknown properties underlying quantum mechanics; they are broadly inspired by de Broglie's pilot-wave approach, seminal for the whole development of quantum physics [53]. In case these experiments would generate reproducible results that contradict quantum mechanics, they could guide the development of new theories. We also argued that contextual and local hidden variables, as probed in the first experiment, are to be expected if one accepts innocuous assumptions, such as that quantum mechanics and general relativity can be unified.

---

[8] There is a vast literature on retrocausality in the context of quantum foundations (e.g. [49, 50, 51, 52]).

These modified Bell experiments are exploratory and do not verify precise numerical predictions of a finalized theory. This situation is reminiscent of particle physics, where most physicists believe that open-ended experiments are necessary for the advancement of the field (see the consensus in e.g. [32]). For instance, even without testing a specific prediction, precision measurements, say of Higgs couplings or neutrino masses, can reveal inconsistencies that guide new theories. In the foundations of quantum mechanics fundamental problems remain that deserve equal attention. Bell experiments allow for highly precise measurements – a key contribution to the precision is the sample size of the photon pairs – but have until now mostly been done within the standard paradigm, focusing on realizing the ideal conditions under which the Bell inequality holds. Modified Bell experiments beyond this paradigm are rare [20, 21, 49, 54]. At the same time such experiments are incomparably less expensive than particle physics experiments. Therefore, we submit that they can offer a platform for advancing the foundations of quantum physics.

[16] J. W. M. Bush, V. Frumkin, and P.J. Sáenz, "Perspectives on pilot-wave hydrodynamics," Appl. Phys. Lett. 125, 030503 (2024).
[17] C. Budroni, A. Cabello, O. Gühne, M. Kleinmann, J.-Å. Larsson, "Kochen-Specker contextuality," arXiv:2102.13036 [quant-ph] (2022).
[18] N. D. Mermin, "Hidden variables and the two theorems of John Bell," Reviews Mod. Physics 65, 803 (1993).
[19] M. Howard, J. Wallman, V. Veitch, & J. Emerson, "Contextuality supplies the 'magic' for quantum computation," Nature, 510(7505), 351-355 (2014).
[20] D. Salart, A. Baas, C. Branciard, N. Gisin, and H. Zbinden, "Testing the speed of 'spooky action at a distance'," Nature 454, 861–864 (2008).
[21] J. Yin, Y. Cao, H.-L.Yong, J.-G. Ren, H. Liang, S.-K. Liao, F. Zhou, C. Liu, Y.-P. Wu, G.-S. Pan, L. Li, N.-L. Liu, Q. Zhang, C.-Z. Peng, and J.-W. Pan, "Lower bound on the speed of nonlocal correlations without locality and measurement choice loopholes," Phys. Rev. Lett. 110, 260407 (2013).
[22] L. de Broglie, "La mécanique ondulatoire et la structure atomique de la matière et du rayonnement," J. Phys. Radium, 8, 225–241 (1927).
[23] P. Jamet, and A. Drezet, "A Time-(Anti) symmetric Approach to the Double Solution Theory," Foundations 5.1: 1 (2024).
[24] J. W. M. Bush. "Pilot-wave hydrodynamics," Ann. Rev. Fluid Mech. 47 (2015).
[25] P. Jamet, and A. Drezet, "A mechanical analog of Bohr's atom based on de Broglie's double-solution approach," Chaos: An Interdisciplinary Journal of Nonlinear Science 31, 10 (2021).
[26] A. Drezet, "A time-symmetric soliton dynamics à la de Broglie," Found. Phys. 53, 72 (2023).
[27] A. Drezet, "Whence nonlocality? Removing spooky action-at-a-distance from the de Broglie–Bohm pilot-wave theory using a time-symmetric version of the de Broglie double solution," Symmetry 16, 8 (2024).
[28] G. 't Hooft, "The Cellular Automaton Interpretation of Quantum Mechanics," (Fundamental Theories of Physics, 185), Springer Open (2016).
[29] G. 't Hooft, "Fast Vacuum Fluctuations and the Emergence of Quantum Mechanics," arXiv:2010.02019 [quant-ph] (2021).
[30] S. Donadi, and S. Hossenfelder, "Toy model for local and deterministic wave-function collapse," Phys. Rev. A 106(2), 022212 (2022).
[31] L. De la Peña, A. M. Cetto, & A. Valdés-Hernández, "The emerging quantum. The Physics Behind Quantum Mechanics," Cham: Springer International Publishing (2015).
[32] J. Cooley, et al., "Report of the topical group on particle dark matter for Snowmass,"arXiv:2209.07426 (2021).
[33] L. Vervoort, "Are Hidden-Variable Theories for Pilot-Wave Systems Possible?," Found. Phys. 48:7, 803-826 (2018).
[34] A. Nachbin, "Walking droplets correlated at a distance", Chaos 28, 096110 (2018).
[35] A. Nachbin, P. A. Milewski, and J. W. M. Bush, "Tunneling with a hydrodynamic pilot-wave model," Phys. Rev. Fluids 2, 034801 (2017).
[36] P. A. Milewski, C. A. Galeano-Rios, A. Nachbin, and J. W. M. Bush, "Faraday pilot-wave dynamics: modelling and computation," J. Fluid Mech. 778, 361 (2015).
[37] K. A. Phillips, R. Cimpeanu, and P. A. Milewski, "Modelling two-dimensional droplet rebound off deep fluid baths," Proc. R. Soc. A 481, 20240956 (2024).
[38] A. Rodriguez, K. Papatryfonos, E. Cardozo de Oliveira, and N. D. Lanzillotti-Kimura, "Topological Nanophononic Interface States Using High-order Bandgaps in the One-Dimensional Su-Schrieffer-Heeger Model," Phys. Rev. B 108, 205301 (2023).
[39] S. Hossenfelder, "Testing super-deterministic hidden variables theories," Found. Phys. 41, 1521 (2011).
[40] G. Feinberg, "Possibility of faster-than-light particles," Phys. Rev. 159, 1089–1105 (1967).
[41] V. Nikolaev, L. Vervoort, "Aspects of Superdeterminism Made Intuitive," Found. Phys. 53, 17 (2023).
16